\documentclass[iop]{emulateapj}

\usepackage{graphicx}
\usepackage{epstopdf}
\usepackage{hyperref}
\usepackage{color}
\newcommand{\msun}{M_{\odot}}
\slugcomment{To Appear in {\rm The Astrophysical Journal Letters}}
\shorttitle{Brightest  cluster - SFR relation in the NIR}
\shortauthors{Z. Randriamanakoto et al.}

\begin{document}


\title{Near-infrared adaptive optics imaging of infrared luminous galaxies:\\ the brightest cluster magnitude - star formation rate relation}


\author{Z. Randriamanakoto$^{1,2}$, A. Escala$^{3}$, P. V{\"a}is{\"a}nen$^{1,4}$, E. Kankare$^{5}$, J. Kotilainen$^{5}$,  S. Mattila$^{5}, $ S. Ryder$^{6}$}
\affil{$^1$South African Astronomical Observatory, P.O. Box 9 Observatory, Cape Town, South Africa}
\affil{$^2$University of Cape Town, Astronomy Department, Private Bag X3, Rondebosch 7701, South Africa}
\email{zara@saao.ac.za}
\affil{$^3$Departamento de Astronom\'ia, Universidad de Chile, Casilla 36-D, Santiago, Chile}
\affil{$^4$Southern African Large Telescope, P.O. Box 9 Observatory, Cape Town, South Africa}
\affil{$^{5}$Finnish Centre for Astronomy with ESO (FINCA), University of Turku, V\"ais\"al\"antie 20, FI-21500 Piikki\"o, Finland}
\affil{$^{6}$Australian Astronomical Observatory, P.O. Box 915, North Ryde, NSW 1670, Australia}

\begin{abstract}

We have established a relation between the brightest super star cluster magnitude in a galaxy and the host star formation rate (SFR) for the first time in the near infrared (NIR).  The data come from a statistical sample of $\sim40$ luminous IR galaxies (LIRGs) and starbursts utilizing $K$-band adaptive optics imaging.  While expanding the observed relation to longer wavelengths, less affected by extinction effects, it also pushes to higher SFRs. The relation we find,  $M_K \sim -2.6 \, \log \rm SFR$, is similar to that derived previously in the optical and at lower SFRs. It does not, however, fit the optical relation with a single optical to NIR color conversion, suggesting systematic extinction and/or age effects.  While the relation is broadly consistent with a size-of-sample explanation, we argue physical reasons for the relation are likely as well. In particular, the scatter in the relation is smaller than expected from pure random sampling strongly suggesting physical constraints.  We also derive a quantifiable relation tying together cluster-internal effects and host SFR properties to possibly explain the observed brightest SSC magnitude vs.\ SFR dependency.

\keywords{galaxies: star formation --- galaxies: star clusters: general --- infrared: galaxies} 
\end{abstract}

\section{Introduction}

Young massive star clusters also known as super star clusters (SSCs) are a good tracer of the current star formation of the host galaxy (\citealp{2010ARA&A..48..431P}). Over the last decade, many studies  have shown that there is an empirical relation between the $V$-band luminosity of the brightest cluster magnitude\footnote{Hereafter, we will refer to the "brightest cluster magnitude" as the "brightest cluster".}  and the global star-formation rate (SFR) of the galaxy (e.g.\ \citealp{2002AJ....124.1393L,  2004MNRAS.350.1503W, 2008MNRAS.390..759B}). Although various reasons have been suggested to explain the relation, they all highlight the important role star clusters play in understanding their host galaxy properties.  \citet{2002AJ....124.1393L} and \citet{2003dhst.symp..153W} emphasize the importance of size-of-sample effect in universal cluster formation, i.e.\ that large SSC populations preferentially sample the initial luminosity functions (LFs) to higher values, while \citet{2004MNRAS.350.1503W} and \citet{2008MNRAS.390..759B}  show through theoretical simulations that  effects from physical processes could also generate the observed correlation.   \citet{2004MNRAS.350.1503W} derive an expression which directly relates the total SFR of the galaxy with the mass of the brightest star cluster,  with the assumption that the most massive cluster is always the brightest. On the other hand, \citet{2008MNRAS.390..759B} argue that the youngest clusters are the brightest implying the tight observed relation is an imprint of the current SFR of the galaxy. \citet{2011MNRAS.417.1904A} find that clusters in Blue Compact Galaxies (BCGs) preferentially lie above the relation fit to more "normal" galaxies suggesting that the environments of SSCs play a role in determining the relation.

So far the validity of the relation has only been tested in the optical regime, and mostly using fairly nearby star-forming galaxies.  Extinction effects are difficult to correct for especially in the dustier galaxies and will necessarily introduce scatter in the relation; using redder wavelengths will significantly improve the situation.  With adaptive optics (AO) it is now possible to probe more distant host galaxies and thus increase the SFR baseline. Will the relation still hold in the near infrared (NIR) and at larger SFR levels? Can we see effects of random sampling, and/or are there clear physical processes behind the relation? In this Letter, we use $K_S$-band\,(AO) imaging of 43 strongly star-forming galaxies  to address these questions.  

%Along with \citet{2011AJ....142...79M} and our pilot study \citet{2013MNRAS.431..554R}, it is now for the first time possible to use statistical samples of luminous infrared galaxies (LIRGs) to study SSC properties. Such targets are interesting because they extend the SSC environments to high and extreme levels of SF activity. 
%We assume the following standard cosmology throughout:  $H_0 = 73$ km s$^{-1}$ Mpc$^{-1}$, $\Omega_M = 0.27$, and $\Omega_{\Lambda} = 0.73$.

\section{Observations and photometry}\label{data}

Most of our sample, 30 galaxies, were imaged in the $K_S$-band with the NACO AO-instrument on the ESO VLT\footnote{Program 086.B-0901 (PI Escala) and 089.D-0847 (PI Mattila)}. 
The targets are IRAS galaxies from the RBGS \citep{2003AJ....126.1607S}, selected to be $D_L \lesssim 200$\,Mpc with IR luminosities above log\,($L_{IR}/L_{\odot}$)~=~10.6, and with a bright enough reference star near the field-of-view required by the AO system.  The galaxies are hence a representative statistical sample of all IR-bright galaxies above log\,($L_{IR}/L_{\odot}$)~=~10.6 within the distance limit. They all have "cool" IRAS colors, though no AGN were excluded {\em a priori}. Depending on the size of the galaxy, either the S27 or S54 camera was used, resulting in a pixel scale of 0.027\,''\,pix$^{-1}$ or 0.054\,''\,pix$^{-1}$, respectively. The final point spread function (PSF) resolution was typically $\sim$\,0.1".  We used a dithering mode with 120\,{\it sec} per pointing, with total integration times per target ranging between 20 and 40 minutes. 

Our {\tt IRAF}-based pipeline was used to perform sky-subtraction. The individual frames were then aligned before average-combining them to get the final science image.  We checked for image quality in individual frames  and those with non-optimal AO-corrections were excluded, resulting in shorter total integration times in some cases (Table\,\ref{coords_table}). Some frames with obviously non-photometric conditions were also excluded.

Photometric zero-points were either retrieved from the ESO/NACO official website, if recorded, 
or estimated by correlating with 2MASS $K_{S}$ point-sources present in the field.  Fields where both methods were available were used to check consistency of photometry.  The zero-point uncertainty of $\sim 0.1$ is included in resulting SSC photometry, which varies in the range $\sim 0.1 - 0.4$ mag. The results are in a Vega-based system.

\begin{table}   
\scriptsize

\caption{\small Observation log of the VLT/NACO and Gemini/NIRI $K$-band data.}
\begin{minipage}{\textwidth}
\begin{tabular}{l@{\hspace{0.10cm}}c@{\hspace{0.10cm}}c@{\hspace{0.10cm}}c@{\hspace{0.10cm}}c@{\hspace{0.10cm}}c@{\hspace{0.10cm}}}   
\hline

Galaxy name &Exp.time &  $D_L$&$M_K^{brightest} $& log\,$L_{\rm IR}$ & SFR  \\
                         &(sec)       &(Mpc) & (mag) & ($L_{\odot}$) &($\rm M_{\odot}yr^{-1}$)  \\
                ~~~~~~(1)    &   (2)        &  (3) & (4) & (5) & (6)  \\         
   
\hline \hline  
\multicolumn{6}{@{} p{8.5 cm} @{}}{\footnotesize{\hspace{3cm} VLT/NACO DATA}}\\\\
ESO\,440-IG058-N &2280 &102.0 &$-15.89$&10.59$^{\dagger}$ &6.6  \\
IC\,2522 &1260 &46.1 &$-15.52$&10.63 &7.3\\
MCG\,-02-01-052 &1860 &110.0 &$-17.21$&10.63$^{\dagger}$ &7.3  \\
NGC\,3620 &1780 &24.9 &$-15.65$&10.70 &8.5\\
ESO\,428-G023 &1770 &44.5 &$-15.51$&10.76 &9.8\\
ESO\,221-IG008 &4620 &46.7 &$-15.90$&10.77 &10.0 \\
IRAS\,$06164+0311$ &1100 &41.5 &$-16.06$&10.79 &10.5\\
NGC\,1134 &1240 &47.4 &$-15.86$&10.83 &11.5 \\
ESO\,491-G020 &1200 &43.5 &$-17.47$&10.86$^{\dagger}$ &12.3\\
NGC\,4433 &1020 &46.3 &$-16.39$&10.87 &12.6\\
NGC\,1204 &1100 &61.4 &$-17.50$&10.88 &12.9  \\
NGC\,3508 &1860 &59.1 &$-16.42$&10.90 &13.5  \\
NGC\,1819 &1950 &61.9 &$-17.11$&10.90 &13.5  \\
MCG\,-02-33-098 &960 &70.8 &$-15.69$&10.95$^{\dagger}$ &15.3  \\
NGC\,4575 &720 &45.0 &$-15.62$&10.96 &15.5\\
NGC\,6000 &2880&32.1&$-16.36$&10.97&15.8\\
ESO\,550-IG025-S &1950 &135.0 &$-16.76$&11.03$^{\dagger}$ &18.2  \\
ESO\,319-G020 &960 &43.2 &$-15.98$&11.04 &18.6 \\
MCG\,+02-02-003 &2340 &70.5 &$-16.19$&11.08 &20.4  \\
ESO\,264-G057 &1120 &75.8 &$-17.15$&11.08 &20.4  \\
ESO\,320-G030 &1120 &49.0 &$-15.70$&11.10 &21.4  \\
ESO\,221-IG010 &240 &45.9 &$-17.19$&11.17 &25.1 \\
ESO\,267-G030 &1920 &80.9 &$-16.86$&11.19 &26.3  \\
ESO\,550-IG025-N &1950 &135.0 &$-17.07$&11.24$^{\dagger}$ &29.5  \\
ESO\,440-IG058-S &2280 &102.0 &$-17.79$&11.28$^{\dagger}$ &32.3 \\
NGC\,3110 &2520 &75.2 &$-17.74$&11.31 &34.7  \\
IRAS\,13052-5711 &3480 &91.6 &$-16.51$&11.34 &37.2  \\
ESO\,264-G036 &2340 &92.0 &$-17.98$&11.35 &38.1  \\
IRAS\,12116-5615 &1800 &117.0 &$-18.42$&11.59 &66.1  \\
IRAS\,F06076-2139 &1560 &160.0 &$-19.61$&11.59 &66.1  \\
IRAS\,01173+1405 &1320 &127.0 &$-18.00$&11.63 &72.5  \\
IRAS\,F01364-1042 &1560 &201.0 &$-20.01$&11.76 &97.8 \\
IRAS\,18293-3413 &1230&74.6&$-18.23$&11.81&109.8 \\
NGC\,6240 &720&103.0&$-18.79$&11.85&120.4\\
IRAS\,19115-2124  &1410&206.0&$-19.71$&11.87&126.0 \\ \hline 
\multicolumn{6}{@{} p{8.5 cm} @{}}{\footnotesize{\hspace{3cm} Gemini/NIRI DATA}}\\\\
IRAS\,F16516-0948 &900&94.8&$-18.10$&11.24&29.5 \\
CGCG\,049-057 &1680&56.4&$-17.14$&11.27&31.6 \\
IRAS\,F17578-0400 &1470&57.3&$-17.56$&11.35&38.1 \\
MCG\,+08-11-002 &1140&79.9&$-17.52$&11.41&43.7 \\
IRAS\,F17138-1017 &990&72.2&$-18.48$&11.42&44.7 \\
NGC\,3690 &2192&45.3&$-17.88$&11.48$^{\dagger}$&51.6 \\
IC\,694 &1260&45.3&$-16.95$&11.66$^{\dagger}$&77.4 \\
IC\,883  &1440&101.0&$-18.38$&11.67&79.5 \\

\hline    
\multicolumn{6}{@{} p{8.5 cm} @{}}{\footnotesize{\textbf{Notes.} Col\,(1): {\tt IRAS} survey name; (2): Total exposure time; (3): Luminosity distance from NED Database; (4):  $K_S$-band absolute magnitude of the brightest cluster; (5): Galaxy IR luminosity from \citet{2003AJ....126.1607S}, any value marked by $\dagger$ is estimated by using the method described in $\S$\,\ref{sec_relation}; (6): SFR derived from Eq.\,\ref{Kennicut}}}
\label{coords_table}
 
\end{tabular}   

\end{minipage}
\end{table}  

We added 12 other LIRGs which form a homogeneous dataset (same wavelength and similar depth and resolution) with the new sample described above.  Eight of these come from Gemini-N/ALTAIR/NIRI and four from earlier NACO data\footnote{Programs 072.D-0433, 073.D-0406, 084.D-0261, 087.D-0444 for earlier NACO data; and GN-2008A-Q- 38, GN-2008B-Q-32, GN-2009A-Q-12, GN-2009B-Q-23, GN-2010A-Q-40 (PI Ryder) for Gemini data.}. The Gemini sample was selected with a higher log\,($L_{IR}/L_{\odot}>$)11.3 cut-off; the final results are checked without these in case of any biases.  Ten of these additional targets were published in \citet{2013MNRAS.431..554R} with the SSC catalogues  ready for analysis, and two unpublished targets are analysed here (NGC 6240 and NGC 6000). We refer the reader to \citet{2007ApJ...659L...9M}, \citet{2008ApJ...689L..97K,2012ApJ...744L..19K}, \citet{2008MNRAS.384..886V, 2008ApJ...689L..37V}, and \citet{2013MNRAS.431..554R} for  details of observations and data reduction.

\subsection{Photometry and SSC selection}\label{Phot}

The SSC candidate photometric catalogues were generated by following the same steps as  in \citet{2013MNRAS.431..554R}. Briefly, an unsharp-masked version of the science image was used for object detection using  {\tt SExtractor} \citep{1996A&AS..117..393B} with a configuration optimized to detect faint sources in very complex backgrounds. Photometry itself was then performed with {\tt IRAF/PHOT} in aperture radii of 2 and 3 pixels (0.11" and 0.08") in frames taken with S54 and S27, respectively. The sky annuli were 1.5 and 2 pixels wide, respectively, with the inner radius one pixel away from the aperture radius in both cases. Aperture correction was achieved with the usual curve-of-growth method, drawn until $\sim$\,1\,'' radius. If there were enough isolated point sources in the field, this aperture correction was dependent on the distance to the AO reference star.  Our SSC selection method retains only likely point-sources by utilizing the concentration of light in the detected sources.  Simultaneously, we exclude likely foreground stars using visual inspection supported by expected star counts from the models of Besan\c{c}on \citep{2003A&A...409..523R}, as well as the nucleus, or nuclei, of the target galaxies.

\section{Results}\label{sec_relation}

Assuming, as usual, that the SFR of strongly star forming galaxies is well represented by their IR luminosity, the following empirical relation by \citet{1998ARA&A..36..189K} was used to convert the $L_{IR}$ to SFR for each galaxy: 
\begin{equation}
{\rm \frac{SFR}{M_{\odot}yr^{-1}} }= 1.7 \times 10^{-10} L_{ IR}\, [L_{\odot}].
\label{Kennicut}
\end{equation}
Both values are listed in Table\,\ref{coords_table}.

If the galaxy is a close pair or a multiple system, its IR luminosity has to be separated into individual components since the \citet{2003AJ....126.1607S} values are for the whole system due to the poor spatial resolution of IRAS. Therefore, we used WISE 12 and 22 $\mu m$, Spitzer/MIPS 24 and 60 $\mu m$, and Herschel/PACS 70 $\mu m$ archival data\footnote{\url{http://irsa.ipac.caltech.edu}, \url{http://herschel.esac.esa.int}}  to measure the flux from each galaxy component and the IRAS-based $L_{IR}$ was redistributed according to the average ratio from all those wavelengths that were available and resolved.
This method was adopted for these targets: ESO\,440-IG058, MCG\,-02-01-052, MCG\,-02-33-098, ESO\,491-IG020, ESO\,550-IG025, IC\,694/NGC\,3690 (=Arp\,299). 

\begin{figure}
\centering
\resizebox{1.0\hsize}{!}{\rotatebox{0}{\includegraphics{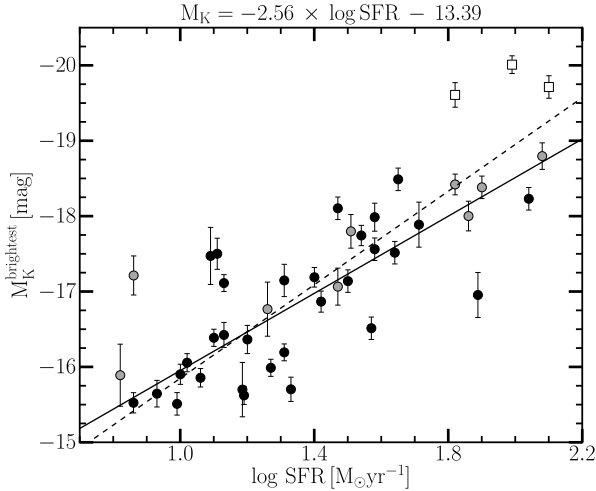}}}
\caption{\small  An empirical relation between the $K$-band magnitude of the brightest cluster and the SFR of the galaxy. The dashed line shows a weighted linear fit to all the data, including the three most distant targets at $D_L > 150\,\rm Mpc$ shown as open squares. The solid line fits the $D_L \leq 150\,\rm Mpc$ targets labeled as circles;  those at $D_L \leq 100\,\rm Mpc$  are black and those at $100 < D_L \leq 150\,\rm Mpc$ are grey.}
\label{relation}
\end{figure}

\begin{figure}
\centering
\resizebox{1.0\hsize}{!}{\rotatebox{0}{\includegraphics{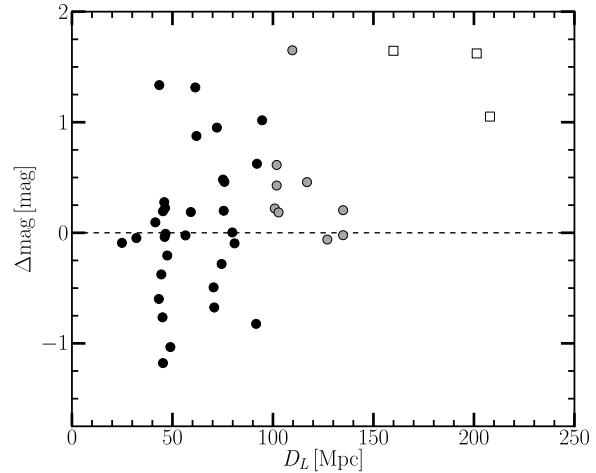}}}
\caption{\small The dispersion about the fit in Fig\,\ref{relation},as a function of distance, from a relation such as shown in Fig.\,\ref{relation} fit for targets closer than 80\,Mpc. Symbols as in Fig.\,\ref{relation}.}
\label{distrelation}
\end{figure}

The NIR brightest cluster - SFR relation is shown in Fig.\,\ref{relation}.  A weighted linear fit to {\em all} the points (the uncertainties of the resolved components' $L_{IR}$ points were doubled) results in the following relation, shown as the dashed line in Fig.\,\ref{relation}:
\begin{equation}
M_K^{brightest} = - 3.10 \times \log \rm\,SFR - 12.75.
\label{slope_all}
\end{equation}
Blending effects may be a concern, however.  The physical spatial resolution in our survey, corresponding to the $\sim0.1$\," PSF size, is typically 20 to 60 pc. Thus, individual detections of SSC candidates could potentially be blends of more than one intrinsic SSC.  The effect may not be too severe for the brightest clusters, since \citet{2013MNRAS.431..554R} showed using Monte-Carlo (MC) simulations that a single bright SSC will overwhelmingly dominate the luminosity of a SSC candidate detection when small apertures are used, except in the most distant targets approaching $D_L \sim 200\,\rm Mpc$, or in case of very strong clustering of SSC regions. Nevertheless, we check for blending effects in the following way:  the $M_K$ vs.\  SFR relation was fit for the "safe" galaxies at $D_L \leq 80$ \,Mpc (the slope is $-2.50$ in this case), and Fig.\,\ref{distrelation} then plots the {\em difference} of the brightest $M_K$ from this best-fit relation vs.\ the distance of the host.  If distance plays no part, a scatter plot is expected. Indeed, no systematics are seen, apart from the 3 most distant targets falling significantly above the null-hypotheses line.  We interpret this as the brightest SSC in those 3 targets potentially being contaminated by other clusters and exclude them from further analysis.  We obtain a new best fit (solid line in Fig.\,\ref{relation}) when using a $D_L \leq150\,\rm Mpc$ constraint:
\begin{equation}
M_K^{brightest} = - 2.56 \times \log \rm\,SFR - 13.39.
\label{slope_150}
\end{equation}
The slope varies in the range $-2.49$ to $-2.56$ with distance limits set in-between 80 and 150\,Mpc.  
The formal uncertainty of the slope is $\pm 0.07$ and the scatter of the observed $M_K$ values is  $\sigma \approx 0.62$\,mag. The slope {\em without} the 8 Gemini galaxies is $-2.43$; the difference is within $\sim1.5\sigma$ of the overall fit and does not affect any conclusions.

\begin{figure}
\centering
\resizebox{1.0\hsize}{!}{\rotatebox{0}{\includegraphics{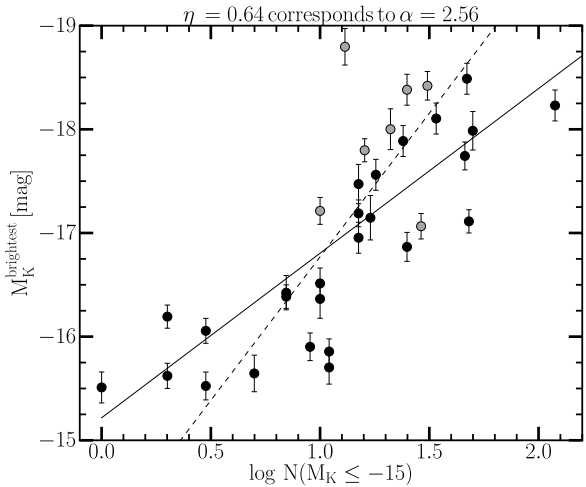}}}
\caption{\small  $M_K^{brightest}$ as a function of the number of SSCs brighter than $-15$ mag for targets with $ D_{L} \leq$ 150\,Mpc. The solid line is the best-fit relation, whereas the dashed line corresponds to a LF power-law slope of $\alpha=1.9$. Symbols as in Fig.\,\ref{relation}.}
\label{N_relation}
\end{figure}

\section{Discussion}

\subsection{Statistical interpretation} \label{statistical}

It has been suggested that the $M^{brightest} - \rm SFR $ relation in the $V$-band can be explained in purely statistical terms (\citealp{2002AJ....124.1393L}). To explore this possibility in the NIR, $M_K^{brightest}$ is plotted against the number of SSC candidates brighter than a certain absolute magnitude level. We select  $M_K = -15\,\rm mag$ since at that level we do not yet need completeness corrections. The result can be seen in Fig.\,\ref{N_relation}. This correlation is consistent with the idea that the more clusters are forming in a galaxy, the higher will be the probability to sample the brightest ones from a given LF of the overall population \citep{2002AJ....124.1393L}.   The empirical relation can also be tied to the slope of the LF.  If the $L^{max}$ of the most luminous object scales with the total number N of the clusters as given by \citet{2003dhst.symp..153W}: 
\begin{equation}
 L^{max} \sim \rm N^{\eta},
 \label{eq_whitmore}
\end{equation}
then by using the equation from \citet{2003AJ....126.1836H}:
\begin{equation}
\eta = \frac{1}{\alpha - 1}
\label{eq_hunter}
\end{equation}
we can derive the power-law slope $\alpha$ of the cluster luminosity function at higher luminosities. The best-fit $\eta = 0.64$ corresponds to $\alpha = 2.56$ which is not unreasonable for bright parts of SSC LFs \citep{2010ARA&A..48..431P}. From the 10 LIRGs in \citet{2013MNRAS.431..554R} we found a slope of $\alpha \sim 1.9$ (flatter than for normal spirals) which corresponds to $\eta = 1.1$  -- this is overplotted in Fig.\,\ref{N_relation} as the dashed line and is seen to represent the data at $\rm log\,N > 0.5$ fairly well.   

The scatter of $\sigma \sim 0.62$\,mag in Fig.\,\ref{relation} is smaller than in relations derived in the optical showing $\sigma \sim 1$\,mag \citep{2002AJ....124.1393L}, unsurprising since extinction effects are smaller. Given that the SFR determination uncertainties should be of the order of 0.4 dex the observed scatter is surprisingly small, however. We ran MC simulations where a given LF with no physical upper limit was sampled purely
randomly and the magnitude of the brightest cluster was recorded.  The $M^{brightest}$ distribution is narrower with steeper LFs parametrized with the power-law index $\alpha$.  In particular, we find $\sigma = 1.15$ for $\alpha= 2.0$,  and $\sigma = 0.77$ and 0.58 for $\alpha=2.5$ and  $3.0$, respectively, each with uncertainties of $\approx 0.02$.  In case of purely statistical sampling the scatter in our relation must hence be the result of steeper than observed LFs of $\alpha > 2.5$ -- or there are other physical characteristics at play which determine the luminosity of the brightest cluster.

In summary, while the characteristics of the number of detected SSCs vs.\ SFR are consistent with a size-of-sample effect, the tightness of the brightest cluster vs.\ SFR relation in particular suggests that it would be premature to reject an underlying physical cause for this relation.

\subsection{Physical interpretation}

Clusters are born of collapsing giant molecular clouds which inevitably are affected by their galactic environments, especially in cases of interacting and merging galaxies -- how this environment exactly defines SSC properties, and disruption, is a matter of intense debate \citep[e.g.][]{2009Ap&SS.324..183L}.  Specifically regarding the brightest cluster vs.\ SFR relation, \citet{2011MNRAS.417.1904A} have shown how SSCs in BCGs appear elevated from the general relation, and suggest this could be a result of a higher cluster formation efficiency in their extreme environments.  \citet{2009A&A...494..539L} and \citet{2009MNRAS.394.2113G} suggest that the characteristic cluster mass may change as a function of environment, and grow in the more intense SFR of interactions and mergers. One appealing possibility for such a change is the lack of large scale rotation in galaxy mergers \citep{2008ApJ...685L..31E,2010ApJ...724.1503W,2013ApJ...763...39E}.

In addition, there might well be internal constraints on SSC properties.  In the following we outline a possible physical interpretation of the  brightness vs.\ SFR relation based on the idea that the total luminosity, and mass, of a stellar cluster is weighted towards its highest mass stars, and these stellar masses may also be correlated with the environments of the clusters (e.g.\,\citealp{2009MNRAS.393..663W}).

The total luminosity of a cluster can be computed for a given initial
mass function (IMF) and mass-luminosity relation. Assuming a mass-luminosity
relation  of the form $L \propto \rm m^{\alpha_{l}}$  and a power law
IMF $\rm (\frac{dN}{dm} \propto m^{-\beta})$, the total luminosity of a cluster is given by
\begin{equation}
L_{tot} \propto \rm M_{cl}^{(\alpha_{l}-\beta+1)\gamma} \, ,
\end{equation}
where $\rm M_{cl}$ is the total mass of the  cluster, which is 
assumed to satisfy a relation with the most massive star of such cluster: $\rm
M_{star}^{max} \sim M_{cl}^{\gamma}$, being  $\rm \gamma \sim
0.45$ estimated from observations and $\rm \gamma \sim 2/3$ predicted
from simulations (see \citealp{2009MNRAS.393..663W} for a review on different estimates). 

Assuming that the most massive unstable gas cloud in a galaxy, $\rm
 M^{max}_{cloud} \sim f^{2}_{gas} M_{gas}$ (\citealp{2008ApJ...685L..31E}),
 leads to the formation of the most massive
SSC ($\rm  M_{cl}  \propto M^{max}_{cloud}$) and taking into account the correlation  between such
a cloud and the SFR in galaxies  $\rm SFR \propto  [M^{max}_{cloud}]^{\small
  \delta}$ with $\rm  \delta \sim 1.5$ \citep{2009arXiv0909.4318E, 2011ApJ...735...56E},  the total luminosity
of the brightest cluster is given by $L_{tot}^{brightest} \propto
\rm SFR^{(\alpha_{l}-\beta+1)\gamma/\delta}$.  Finally, this can be expressed in terms of absolute
magnitude by $M_{K}^{brightest} \propto - 2.5 \,{\rm \log} \,
L_{K}^{brightest}$, resulting in:
\begin{equation}
M_{K}^{brightest} \propto -2.5  \rm \, \, \frac{(\alpha_{l}-\beta+1)\gamma}{\delta} \,\,  \log\,SFR \, .
\label{eqtheo}
\end{equation}

For a Salpeter IMF ($\rm \beta$ = 2.35), $\rm \delta$ = 1.5
\citep{2009arXiv0909.4318E, 2011ApJ...735...56E}, $\rm \gamma \sim 0.45$ estimated from
observations \citep{2009MNRAS.393..663W} and  a slope of the
mass-luminosity relation of $\rm \alpha_{l} \sim 5$, Eq.\,\ref{eqtheo}
gives a slope  closer to $-3$ in the brightest cluster-SFR relation, which is
comparable to the slope observed if we use the whole sample (Eq.\,\ref{slope_all}). On the other hand, for  a the slope of the
mass-luminosity relation $\rm \alpha_{l}$ of 4, Eq. $\ref{eqtheo}$  gives a slope 
$\rm\sim -2$, which is closer to that derived from our data excluding potentially blended cases (Eq.\,\ref{slope_150}). 

Unfortunately, we do not have a good estimate for the slope of the mass-luminosity relation $\rm \alpha_{l}$ at high masses -- it may, for example, vary in between 1.76 and 8.87 depending on the highest mass of a  star in a cluster for masses larger than $\rm 7\,\msun$ \citep{2003ApJ...584..797P}. Nevertheless, Eq.\,\ref{eqtheo} may be used as a simple physical interpretation of the effects characterizing SSC properties at scales ranging from internal to galactic.

\subsection{Comparison to the $V$-band relation}

\begin{figure}
\centering
\resizebox{1.0\hsize}{!}{\rotatebox{0}{\includegraphics{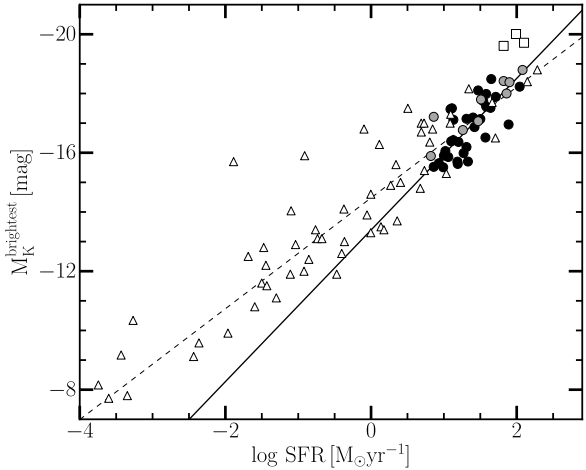}}}
\caption{\small The $M^{brightest} - \,\rm SFR $ relation with data  from literature (the triangles, \citet{2011MNRAS.417.1904A} and references therein) added to the present work (symbols as in Fig.\,\ref{relation}). The solid line is our best fit of Eq.\,\ref{slope_150} and the dashed line is the fit from \citet{2004MNRAS.350.1503W} to the optical $V$-band data after a constant $V-K = 2$ conversion. }
\label{bastian}
\end{figure}

Figure~\ref{bastian} shows the brightest cluster - SFR relation with an expanded scale.  The triangles are $V$-band data as compiled by  \citet{2011MNRAS.417.1904A}, assuming a constant $V-K = 2$ conversion, typical for a $\sim 10$\,Myr age stellar population.  The solid line is the best fit slope of $-2.56$ from our own NIR data extrapolated to lower SFRs.  It appears that the optical points would require a slightly flatter slope, and indeed \citet{2004MNRAS.350.1503W} find $\sim -1.9$ shown as the dashed line, though \citet{2002AJ....124.1393L}  derives $\sim -2.5$ from a subset of the data.  The simplest explanation could perhaps be extinction:  the highest SFR galaxies, LIRGs and ULIRGs, are predominantly interactions and mergers with more dust  on average than lower SFR galaxies \citep{2013A&A...553A..85P}.  The uncorrected optical points at higher SFR could lie too low artificially, thus flattening the slope.  However, with this data-set alone it is not possible to confirm this -- it could as well be that the brightest clusters detected in $V$-band are not necessarily (always) the most luminous clusters in $K$-band possibly implying age effects.  That the $\it slope$ appears slightly different necessarily points to some systematic effects along the SFR base-line,  i.e.\  a constant $V-K$ shift is not appropriate.   

The slope may become steeper if the points at higher SFR, preferentially, are bound results of mergers of individual SSCs in very dense star cluster complexes (\citealp{ 2005ApJ...630..879F}). We also note that outliers {\em below} the relation can be understood as cases where the brightest SSC is not detected, or is severely extincted. And as discussed by \citet{2008MNRAS.390..759B}, outliers {\em above} the line may be cases where the detected cluster is significantly {\em older} than the general population of brightest SSCs.  These are very interesting questions to tackle with combinations of optical and NIR data in the future.

\section{Summary and future work}

From a $K$-band AO sample of 43 strongly star-forming galaxies, mostly LIRGs, we establish the brightest cluster magnitude - SFR relation in the NIR regime. The relation is much less affected by extinction effects than similar comparisons in the $V$-band.   We find a slope  of $-2.56$ which is similar to those from optical derivations made at lower star formation rate levels, though the extension of our slope appears to not be consistent with the full range of optical SSC luminosities if a single $V-K$ conversion is adopted.  We suggest that a systematic extinction effect, where SSCs in higher SFR hosts live in dustier environments, would be a simple explanation for the trend, but systematic age differences may also be involved.

A good correlation of the most luminous cluster and the number of SSCs with $M_{K}$ magnitude brighter than $-15$ shows that a size-of-sample effect is broadly consistent with the observed $M_{K}^{brightest}-\rm SFR$\, relation. On the other hand, the observed scatter in the relation is surprisingly small, and we show that it can be explained with random sampling effects {\em only} if the LF of SSCs is very steep at the bright end, steeper than usually observed.  Hence, physical reasons determining the luminosity of the brightest SSC from host properties, and/or internal cluster effects, likely play a role as well. We derived a relation tying the stellar IMF and mass-luminosity relations together with the global SF properties of the host in explaining the observed brightest cluster magnitude - SFR relation.

In the next steps of the work we will investigate in more detail the environments and extinctions of the host galaxies, and masses and ages of the SSCs inside them, with a combination of optical and NIR data and kinematic information.  These will allow more secure disentanglement of the various physical effects governing the lives and characteristics of super star clusters in galaxies.

 \acknowledgements
We thank the anonymous referee for the valuable comments and suggestions to improve this work. ZR acknowledges funding from the South African Square Kilometre Array, AE from the Financiamiento Basal Grant PFB 06, FONDECYT Grant 1130458 and Anillo de Ciencia y Tecnologia Grant ACT1101, and PV from the National Research Foundation.

%%%%%%%%%%%%%% REFERENCES %%%%%%%%%%%%%%%%%%%%%%%%%%%%%%%%%%%%%
%

%
%
%%%%%%%%%%%%%%%%% FIGURES %%%%%%%%%%%%%%%%%%%%%%%%%%%%%%%%%%%%%


\begin{thebibliography}{27}
\expandafter\ifx\csname natexlab\endcsname\relax\def\natexlab#1{#1}\fi

\bibitem[{{Adamo}, {{\"O}stlin} \& {Zackrisson}(2011)}]{2011MNRAS.417.1904A} 
{Adamo}, A., {{\"O}stlin}, G. \& {Zackrisson}, E. 2011, \mnras, 417, 1904

\bibitem[{{Bastian}(2008)}]{2008MNRAS.390..759B}
{Bastian}, N. 2008, \mnras, 390, 759

\bibitem[{{Bertin} \& {Arnouts}(1996)}]{1996A&AS..117..393B}
{Bertin}, E. \& {Arnouts}, S. 1996, \aaps, 117, 393

\bibitem[{{Escala}(2009)}]{2009arXiv0909.4318E}
{Escala}, A. 2009, ArXiv e-prints

\bibitem[{{Escala}(2011)}]{2011ApJ...735...56E}
---. 2011, \apj, 735, 56

\bibitem[{{Escala} {et~al.}(2013){Escala}, {Becerra}, {del Valle}, \&
  {Castillo}}]{2013ApJ...763...39E}
{Escala}, A., {Becerra}, F., {del Valle}, L., \& {Castillo}, E. 2013, \apj,
  763, 39

\bibitem[{{Escala} \& {Larson}(2008)}]{2008ApJ...685L..31E}
{Escala}, A. \& {Larson}, R.~B. 2008, \apjl, 685, L31

\bibitem[{{Fellhauer} \& {Kroupa}(2005)}]{2005ApJ...630..879F}
{Fellhauer}, M. \& {Kroupa}, P. 2005, \apj, 630, 879

\bibitem[{{Gieles}(2009)}]{2009MNRAS.394.2113G}
{Gieles}, M. 2009, \mnras, 394, 2113

\bibitem[{{Hunter} {et~al.}(2003){Hunter}, {Elmegreen}, {Dupuy}, \&
  {Mortonson}}]{2003AJ....126.1836H}
{Hunter}, D.~A., {Elmegreen}, B.~G., {Dupuy}, T.~J., \& {Mortonson}, M. 2003,
  \aj, 126, 1836
  
 \bibitem[{{Kankare} {et~al.}(2008){Kankare}, {Mattila}, {Ryder},
  {P{\'e}rez-Torres}, {Alberdi}, {Romero-Canizales}, {D{\'{\i}}az-Santos},
  {V{\"a}is{\"a}nen}, {Efstathiou}, {Alonso-Herrero}, {Colina}, \&
  {Kotilainen}}]{2008ApJ...689L..97K}
{Kankare}, E., {Mattila}, S., {Ryder}, S., {P{\'e}rez-Torres}, M., {Alberdi},
  A., {Romero-Canizales}, C., {D{\'{\i}}az-Santos}, T., {V{\"a}is{\"a}nen}, P.,
  {Efstathiou}, A., {Alonso-Herrero}, A., {Colina}, L., \& {Kotilainen}, J.
  2008, \apjl, 689, L97

\bibitem[{{Kankare} {et~al.}(2012){Kankare}, {Mattila}, {Ryder},
  {V{\"a}is{\"a}nen}, {Alberdi}, {Alonso-Herrero}, {Colina}, {Efstathiou},
  {Kotilainen}, {Melinder}, {P{\'e}rez-Torres}, {Romero-Ca{\~n}izales}, \&
  {Takalo}}]{2012ApJ...744L..19K}
{Kankare}, E., {Mattila}, S., {Ryder}, S., {V{\"a}is{\"a}nen}, P., {Alberdi},
  A., {Alonso-Herrero}, A., {Colina}, L., {Efstathiou}, A., {Kotilainen}, J.,
  {Melinder}, J., {P{\'e}rez-Torres}, M.-A., {Romero-Ca{\~n}izales}, C., \&
  {Takalo}, A. 2012, \apjl, 744, L19
  
  \bibitem[{{Kennicutt}(1998)}]{1998ARA&A..36..189K}
{Kennicutt}, Jr., R.~C. 1998, \araa, 36, 189
  
  \bibitem[{{Lamers}(2009)}]{2009Ap&SS.324..183L}
{Lamers}, H.~J.~G.~L.~M. 2009, \apss, 324, 183

\bibitem[{{Larsen}(2002)}]{2002AJ....124.1393L}
{Larsen}, S.~S. 2002, \aj, 124, 1393

\bibitem[{{Larsen}(2009)}]{2009A&A...494..539L}
---. 2009, \aap, 494, 539

\bibitem[{{Mattila} {et~al.}(2007){Mattila}, {V{\"a}is{\"a}nen}, {Farrah},
  {Efstathiou}, {Meikle}, {Dahlen}, {Fransson}, {Lira}, {Lundqvist},
  {{\"O}stlin}, {Ryder}, \& {Sollerman}}]{2007ApJ...659L...9M}
{Mattila}, S., {V{\"a}is{\"a}nen}, P., {Farrah}, D., {Efstathiou}, A.,
  {Meikle}, W.~P.~S., {Dahlen}, T., {Fransson}, C., {Lira}, P., {Lundqvist},
  P., {{\"O}stlin}, G., {Ryder}, S., \& {Sollerman}, J. 2007, \apjl, 659, L9

\bibitem[{{Parravano} {et~al.}(2003){Parravano}, {Hollenbach}, \&
  {McKee}}]{2003ApJ...584..797P}
{Parravano}, A., {Hollenbach}, D.~J., \& {McKee}, C.~F. 2003, \apj, 584, 797

\bibitem[{{Piqueras L{\'o}pez}{et~al.}(2013){Piqueras L{\'o}pez}, J., {Colina}, L., {Arribas}, S. and 
	{Alonso-Herrero}, A.}]{2013A&A...553A..85P}{Piqueras L{\'o}pez}, J., {Colina}, L., {Arribas}, S. \& 
	{Alonso-Herrero}, A. 2013, A\&A, 553A, 85

\bibitem[{{Portegies Zwart}, {McMillan}, \& {Gieles}(2010){Portegies Zwart}, S.~F., {McMillan}, S.~L.~W. \& {Gieles}, M.}]{2010ARA&A..48..431P}{Portegies Zwart}, S.~F., {McMillan}, S.~L.~W. \& {Gieles}, M. 2010, \araa, 48, 431

\bibitem[{{Randriamanakoto} {et~al.}(2013){Randriamanakoto},
  {V{\"a}is{\"a}nen}, {Ryder}, {Kankare}, {Kotilainen}, \&
  {Mattila}}]{2013MNRAS.431..554R}
{Randriamanakoto}, Z., {V{\"a}is{\"a}nen}, P., {Ryder}, S., {Kankare}, E.,
  {Kotilainen}, J., \& {Mattila}, S. 2013, \mnras, 431, 554

\bibitem[{{Robin} {et~al.}(2003){Robin}, {Reyl{\'e}}, {Derri{\`e}re}, \&
  {Picaud}}]{2003A&A...409..523R}
{Robin}, A.~C., {Reyl{\'e}}, C., {Derri{\`e}re}, S., \& {Picaud}, S. 2003,
  \aap, 409, 523

\bibitem[{{Sanders} {et~al.}(2003){Sanders}, {Mazzarella}, {Kim}, {Surace}, \&
  {Soifer}}]{2003AJ....126.1607S}
{Sanders}, D.~B., {Mazzarella}, J.~M., {Kim}, D., {Surace}, J.~A., \& {Soifer},
  B.~T. 2003, \aj, 126, 1607

\bibitem[{{V{\"a}is{\"a}nen} {et~al.}(2008{\natexlab{a}}){V{\"a}is{\"a}nen},
  {Mattila}, {Kniazev}, {Adamo}, {et~al.}}]{2008MNRAS.384..886V}
{V{\"a}is{\"a}nen}, P., {Mattila}, S., {Kniazev}, A., {Adamo}, A., {et~al.}
  2008{\natexlab{a}}, \mnras, 384, 886

\bibitem[{{V{\"a}is{\"a}nen} {et~al.}(2008{\natexlab{b}}){V{\"a}is{\"a}nen},
  {Ryder}, {Mattila}, \& {Kotilainen}}]{2008ApJ...689L..37V}
{V{\"a}is{\"a}nen}, P., {Ryder}, S., {Mattila}, S., \& {Kotilainen}, J.
  2008{\natexlab{b}}, \apjl, 689, L37
  
  \bibitem[{{Weidner} {et~al.}(2010){Weidner}, {Bonnell}, \&
  {Zinnecker}}]{2010ApJ...724.1503W}
{Weidner}, C., {Bonnell}, I.~A., \& {Zinnecker}, H. 2010, \apj, 724, 1503

\bibitem[{{Weidner} {et~al.}(2004){Weidner}, {Kroupa}, \&
  {Larsen}}]{2004MNRAS.350.1503W}
{Weidner}, C., {Kroupa}, P., \& {Larsen}, S.~S. 2004, \mnras, 350, 1503

\bibitem[{{Weidner} {et~al.}(2009){Weidner}, {Kroupa}, \&
  {Maschberger}}]{2009MNRAS.393..663W}
{Weidner}, C., {Kroupa}, P., \& {Maschberger}, T. 2009, \mnras, 393, 663

\bibitem[{{Whitmore}(2003)}]{2003dhst.symp..153W}
{Whitmore}, B.~C. 2003, in A Decade of Hubble Space Telescope Science, ed.
  {M.~Livio, K.~Noll, \& M.~Stiavelli}, 153--178

\end{thebibliography}
\end{document}